\documentstyle[aps,preprint]{revtex}
\tightenlines
\begin{document}
\title{Dbrane boundstate wavefunctions}
\author{Vipul Periwal}
\address{Department of Physics,
Princeton University,
Princeton, New Jersey 08544}

\def\dd{\hbox{d}}
\def\tr{\hbox{tr}}\def\Tr{\hbox{Tr}}
\maketitle
\begin{abstract}  A simple WKB approximation gives explicit 
information about D0brane 
boundstate wavefunctions, suggesting that at large $N$ 
each individual D0brane has a wavefunction $\exp(- c 
r^{9/2}N^{-1/2}).$  Thus the velocity dependent interaction energy 
$v^{4}r^{-7}$
leads to an effective confining potential that grows as $r^{7}.$
\end{abstract} 

\def\al{\alpha}
\def\be{\beta}
\def\la{\lambda}
\def\eps{\epsilon}
\def\sig{\sigma}
\def\la{\lambda}
\def\ga{\gamma}
\def\half{\hbox{$1\over 2$}}
\def\quart{\hbox{$1\over 4$}}
\def\ee{\hbox{e}}
\def\part{\partial}
\def\refe#1{eq.~(\ref{#1})\ }
Dbrane boundstate wavefunctions play a prominent r\^ole in 
recent advances in non-perturbative string theory\cite{bfss}.   
The form of these wavefunctions seems to be quite elusive, though the
existence of the boundstates has been proved in some cases 
cases\cite{sethi,others}.  
Polchinski has obtained results about the boundstate wavefunctions
by other methods\cite{joe}.

I show in this note that a standard physics calculation gives a 
surprisingly simple intuitive form for the wavefunction.  The $N$
dependence of this wavefunction is automatically consistent with 
holography, as pointed out in section 7 of \cite{bfss} by a scaling 
argument, but the 
explicit form of the wavefunction we will find below appears to be new.
Important earlier work that is relevant background for the simple analysis 
presented here is that of Danielsson, Ferretti and Sundborg\cite{dfs}, 
and Kabat and Pouliot\cite{kp}.  These references studied 
the problem by quite different methods. While 
there is some similarity to
what we will find, the form of the wavefunction found here does not 
appear in these papers\cite{dfs,kp}.

I will consider only D0branes in the following.  $N$ static D0branes 
preserve half the supersymmetries of    type IIA string theory
so their energy is independent of their relative positions.  Consider 
a configuration of $N$ D0branes at position $0$ and $1$ 
D0brane  at position $r$ along the $x^{1}$ axis.  We want to compute the
matrix element
\begin{equation}
	\langle 
	N,x^{1}=0;1,x^{1}=r|\exp(-HT)|N+1,x^{1}=0 \rangle\ 
	\label{matrix}
\end{equation}
where $H$ is the Hamiltonian of the system.
We expect on elementary  grounds that for large $T$ this matrix element
should be dominated by the state(s) $|\psi_{0}\rangle$
in the Hilbert space with vanishing
energy, giving us some insight into the
overlap $\langle \psi_{0}|N +1,x^{1}=0 \rangle$ relative to
the overlap $\langle \psi_{0}|N,x^{1}=0;1,x^{1}=r\rangle^{*}.$  

On the other hand, we can compute \refe{matrix} by using Euclidean
functional integrals, evaluated at saddle-point trajectories that 
take us from one point in the classical configuration space 
$C_{1}\equiv (N,x^{1}=0;N_{1},x^{1}=r)$ to another point 
$C_{2}\equiv (N+1,x^{1}=0;N_{1}-1,x^{1}=r).$
Both these configurations have the same classical energy, so this
problem is similar to familiar `tunneling' calculations.  Of course, 
there is a crucial  difference in that there is actually no `potential' 
energy in the system.  The only energy of interaction that appears in 
the system is when the D0branes have non-zero relative velocities, but 
since there must be some non-vanishing relative velocity for motion
from $C_{1}$ to $C_{2},$ this interaction energy is non-vanishing and
leads to a simple explicit result.  As I shall show, the 
velocity-dependent interaction energy leads effectively to a 
wavefunction similar to that of a particle moving in a potential
$\propto r^{7}.$  Thus the potential is much flatter than that of a
harmonic oscillator at small values of $r$ and much steeper at
larger values of $r.$

The action of a test D0brane 
in the configuration $C_{1}$ takes the form
\begin{equation}
	S \approx \int dt\ \left[{1\over 2g} {\dot x}^{2} + {15N \over 16}
	 {{\dot x}^{4}\over 
	r^{7}} 
	  \right]\equiv \int dt\ \left[ {1\over 2g} {\dot x}^{2} - V 
	\right],
	\label{action}
\end{equation}
where $r $  is the distance of the test D0brane from the   cluster  of 
D0branes.  This action is only valid when $r $ is  large enough, and 
velocities small enough, 
but we are interested in   the Euclidean action which has the same form with 
$15/16\rightarrow -15/16.$  The Euclidean motion then corresponds to 
a {\it repulsion} from the  cluster  of D0branes, so the velocity of the
test D0brane vanishes as it approaches the cluster.  It is a 
non-perturbative consequence of supersymmetry 
that the force must vanish in the limit of vanishing velocity, so 
we may be able to trust the action for the Euclidean motions of 
interest even though we
cannot trust it for general Minkowski motions.  This is nevertheless a point 
that needs to be carefully considered since the supersymmetry itself is not
easily continued to Euclidean signature.
The conserved quantity  associated with
Euclidean motions is $\epsilon_{E}= {1\over 2g}{\dot x}^{2}+3V_{E}.$  
The Euclidean
action for a  solution of the equations of motion with the desired 
boundary conditions is then 
\begin{equation}
S_{E}=	T\epsilon_{E}  + 2\int V_{E} \ dt .
\end{equation}
$\epsilon_{E}=0$ for the motion with the smallest Euclidean action, so
we see that
\begin{equation}
	{\dot x}^{2} = {45N \over 8}
	 {{\dot x}^{4} \over r^{7}} 
\end{equation}
which in particular shows that as the test D0brane approaches the 
cluster of zero-branes its velocity decreases as $r^{7/2}.$  Finally
\begin{equation}
	S_{E} = \int_{0}^{r} dx\ {2\sqrt{2}\over9\sqrt{15gN}} x^{{7/2}}dx \propto 
	r^{{9/2}} N^{-1/2}.
\end{equation}
The determinants for fluctuations about this solution should (mostly) 
cancel 
due to supersymmetry, so we are left with
\begin{equation}
\langle \psi_{0}|N,x^{1}=0;1,x^{1}=r\rangle \propto \exp\left(- c r^{{9/2}} 
N^{-1/2}\right)\ .
\label{wavefunction}
\end{equation}

The coefficient $c$ in \refe{wavefunction} depends on   higher
order terms which could be  included in \refe{action}  but 
the important point is that for large $N$ the explicit wavefunction shows 
that the test D0brane is essentially confined to a flat box  of size
$N^{1/9}.$ This power of $N$ is
as expected from the discussion in section 7 of \cite{bfss}.  
However, 
the power of $r$ in \refe{wavefunction} was not explicitly 
computed in \cite{bfss}---it is amusing to find that a velocity 
dependent interaction energy that falls off as $v^{4}r^{-7}$ leads to a
confining potential that grows as $r^{7}.$  The picture is therefore
in accord with a bag model of gravitons with
D0branes as constituents, the parton intuition given in \cite{bfss}.
The elementary analysis presented here is somewhat different from the
sophistication of other approaches\cite{sethi,others}, but it is 
hoped that some intuition into the structure of Dbrane bound states 
can be gained from extensions of this calculation.

\acknowledgements
This work was supported in part by NSF grant PHY96-00258.  I am very 
grateful to I. Klebanov, G. Lifschytz and R.C. Myers for helpful
conversations, and to M. Gutperle for pointing out important 
references\cite{dfs,kp}.

\end{document}